\documentstyle[preprint,eqsecnum,aps,pra]{revtex}

\begin{document}

\renewcommand{\thefootnote}{\fnsymbol{footnote}}
\draft
\title{ Pattern selection in the absolutely unstable regime \\
 as a nonlinear eigenvalue problem:  Taylor vortices in axial flow}
\author{P.~B\"{u}chel, M.~L\"{u}cke, D.~Roth\\}
\address{Institut f\"{u}r Theoretische Physik, Universit\"{a}t
des Saarlandes, D-66041~Saarbr\"{u}cken, Germany\\}
\author{R.~Schmitz\\}
\address{Institut f\"{u}r Festk\"{o}rperforschung, Forschungszentrum,
         D-52425~J\"{u}lich, Germany\\}
\renewcommand{\thefootnote}{\arabic{footnote}}
\setcounter{footnote}{0}

\maketitle

%%%%%%%%%%%%%%%%%%%%%%%%%%%%%%%%%%%%%%%%%%%%%%%%%%%%%%%%%%%%%%%%%%%%%%%
%
%  ABSTRACT
%
%%%%%%%%%%%%%%%%%%%%%%%%%%%%%%%%%%%%%%%%%%%%%%%%%%%%%%%%%%%%%%%%%%%%%%%

\begin{abstract}

A unique pattern selection in the absolutely unstable regime of a
driven, nonlinear, open-flow system is analyzed: The spatiotemporal
structures of rotationally symmetric vortices that propagate 
downstream in the annulus of the rotating Taylor-Couette system due 
to an externally imposed axial through-flow are investigated for 
two different axial boundary conditions at the in- and outlet.
Detailed quantitative results for the oscillation frequency, the axial 
profile of the wave number, and the temporal Fourier amplitudes of 
the propagating vortex patterns obtained by numerical simulations of 
the Navier-Stokes equations are compared with results of the 
appropriate Ginzburg-Landau amplitude equation approximation and also 
with experiments. Unlike the stationary patterns in systems without 
through-flow the spatiotemporal structures of propagating vortices 
are independent of parameter history, initial conditions, 
and system's length. They do, however, depend on the
axial boundary conditions in addition to the driving rate of the inner
cylinder and the through-flow rate. Our analysis of the amplitude 
equation shows that the pattern selection can be described by a 
nonlinear eigenvalue problem with the frequency being the eigenvalue.
The complex amplitude being the corresponding
eigenfunction describes the axial structure of intensity and wave 
number. Small, but characteristic differences in the
structural dynamics between the Navier-Stokes equations and the 
amplitude equation
are mainly due to the different dispersion relations. Approaching the
border between absolute and convective instability the eigenvalue 
problem becomes effectively linear and the selection mechanism 
approaches that one of linear front propagation.

\end{abstract}


% Beginn of the text

\narrowtext

%%%%%%%%%%%%%%%%%%%%%%%%%%%%%%%%%%%%%%%%%%%%%%%%%%%%%%%%%%%%%%%%%%%%%%%
%
%  SECTION I: Introduction
%
%%%%%%%%%%%%%%%%%%%%%%%%%%%%%%%%%%%%%%%%%%%%%%%%%%%%%%%%%%%%%%%%%%%%%%%

\section{Introduction}

In many nonlinear continuous systems dissipative structures branch 
out of a homogeneous basic state when the external stress exceeds 
a critical threshold. Examples for these 
transitions are Taylor-Couette flow, Rayleigh-B\'enard convection, 
binary-fluid convection, flame-front propagation, and some chemical 
or biological processes \cite{CH-RevModPhys-93}. Often,
for a fixed configuration of parameters and boundary conditions a
continuous or discrete family of patterns with different wave numbers 
is stable.  Their stability regime, e.~g., a band of wave numbers 
might be limited by the possibility of resonant triad interactions 
of modes like those described by the 
Eckhaus or Benjamin-Feir mechanism  \cite{CH-RevModPhys-93}.  
The stable structures
within such a band can be generated by appropriately engineered time
histories of the parameters and/or by properly changing the boundary
conditions, e.~g., the system size. The most intensively investigated
examples in this respect are the structures of Taylor vortices
\cite{CH-RevModPhys-93,ACDH-PhysD-86,AC-PhysLett-83,HAC-PhysRev-86,%
LMKW-Springer-87,LMW-PhysRevA-85,RP-PhysLett-87}
in an annulus between concentric cylinders of which the inner one 
rotates and convective roll patterns in horizontal layers of 
one-component fluids
\cite{CH-RevModPhys-93,Croquette,RBWB-PhysLett-87,FS86,LMK-PhysRevA-87,%
LMKW-Springer-87}
or binary mixtures heated from below \cite{K-PhysA-92,BES-PhysA-92}.

This multiplicity of solutions of the underlying nonlinear partial
differential equations that stably coexist for a fixed configuration of
parameters and boundary conditions seems to disappear in an open-flow
system: Recent numerical simulations of Rayleigh-B\'enard convective 
rolls traveling downstream in an imposed horizontal Poiseuille flow  
showed
\cite{MLK-Euro-89,MLK-PhysRevA-92,MLK-Nato-92} that their structure is
uniquely selected -- i.~e. it is  independent of parameter history, 
initial conditions, and system size -- in the absolutely unstable 
regime.  This is the parameter regime of an open-flow system in 
which the secondary
pattern starting, e.~g., from a spatially localized perturbation can 
grow in upstream as well as in downstream direction 
\cite{BersBriggs}.
By contrast, in the convectively unstable regime 
\cite{Huerre} initial perturbations
are blown out of the system -- both, the  upstream as well as the
downstream facing front of the growing structure move downstream. In 
the absolutely  unstable regime the structure expansion proceeds  until 
the upstream (downstream) moving front  encounters in a finite system 
the inlet (outlet) and adjusts to the inlet (outlet) boundary 
condition. The final
pattern resulting in such a situation shows a characteristic streamwise
profile of the amplitude growing with increasing distance 
from the inlet and of the wave number variation, and a characteristic 
global oscillation frequency associated with the downstream 
motion of the pattern.

In this work we elucidate in numerical and analytical detail how such a
uniquely selected spatiotemporal pattern structure can be understood 
as a
nonlinear  eigenvalue problem with the oscillation frequency being the
eigenvalue and the profiles of pattern intensity and wave number
determining the corresponding eigenfunction. We also show how this 
pattern
selection process is related to the one occurring behind a front or 
"domain wall" that spatially separates an unstable, homogeneous state 
from a stable, structured state. To that end we present results of 
extensive  numerical
and analytical investigations of vortex patterns in the  annulus of the
Taylor-Couette setup with an externally imposed axial through-flow.  
For small through-flow
rates and small rotation rates of the inner cylinder 
the structure of propagating vortex~(PV) flow with vortices being 
advected in downstream direction is rotationally symmetric.
Only for higher through-flow rates, which are not investigated
here, there is a bifurcation \cite{CA-FluidMech-77,TJ-FluidMech-81} to
spirals
\cite{TJ-FluidMech-81,S-Lond-61,S-Ann-65,GuFahidy,GF-Can-86,TS3-93,%
LDM-PhysFluids-92}.
We have performed numerical simulations of the full, rotationally
symmetric, two dimensional~(2D)~Navier-Stokes equations~(NSE). 
They are compared with our
numerical and analytical results obtained from the appropriate
\cite{RLM-PhysRevE-93} 1D~Ginzburg-Landau amplitude equation~(GLE)
approximation to the problem of PV~flow and with experimental results
\cite{TJ-FluidMech-81,S-Lond-61,S-Ann-65,TS3-93,TS-Euro-91,TS1-93}

The eigenvalue problem of the pattern selection by the imposed flow can
best be analyzed and explained within the GLE~framework. Therein, the
complex amplitude $A(z,t)$ of the PV~flow depends  in the absolutely 
unstable regime on the streamwise position $z$ and on
time $t$ in a multiplicative way only, 
\begin{equation} A(z,t) = a(z) e^{-i \Omega t},\end{equation}
after transients have died out.
The $z$-independent frequency $\Omega$ and the
complex, $z$-dependent amplitude $a(z)$ that describes the streamwise
variation of pattern modulus and wave number are simultaneously fixed 
via
a solvability condition in the form of a nonlinear eigenvalue problem. 
The selection of $\Omega$ and $a(z)$ seems to result from requiring 
the spatial
variation of the amplitude to be as small as possible under the imposed 
boundary conditions for $A$ at the ends of the annulus. 

Approaching the border between the absolutely and convectively unstable
regimes the pattern selection mechanism becomes linear: For driving and
through-flow rates on this border line the selected frequency is the one
resulting from a linear front whose spatiotemporal behavior is governed 
by the fastest growing linear mode. The latter is identified by 
a particular
saddle of the complex linear dispersion relation over the complex wave
number plane \cite{CH-RevModPhys-93}.  Now, the linear dispersion 
relations of NSE \cite{RDprivate} and GLE differ for supercritical 
control
parameters.  And  therefore the PV~structures selected by NSE or GLE 
differ in a characteristic way.

It should be emphasized that the structural dynamics of pattern 
formation
in the convectively unstable regime at larger through-flow rates and/or
smaller driving rates substantially differs from the one investigated 
here
in the absolutely unstable regime. The latter regime is governed by
nonlinear contributions in the balance equations; the resulting  
patterns
are uniquely selected and insensitive to initial conditions, parameter
history, and small perturbations. On the other hand, in the convectively
unstable regime that has attracted more experimental activities lately
\cite{TS-Euro-91,TS1-93,BAC-PhysRev-91,BCA-PhysD-92,BAC-Pre-93,%
TS-PhysRev-91,TGS2-93}
the growing patterns are sensitive to initial conditions and 
perturbations. Thus, e.~g., the noise sustained patterns \cite{Deissler}
occurring in this regime depend on details of the spatiotemporal 
properties of the perturbation source. 

The Taylor-Couette system \cite{DIPS-Springer-81} with an imposed
axial  through-flow has been investigated as a well defined open-flow
system theoretically  \cite{G-ProcPhilos-37} and experimentally
\cite{CornishFage}  since the beginning 1930s. Linear
stability analyses of the basic state to
traveling axisymmetric vortices were performed using various 
approximations, e.~g.~, 
for the narrow-gap limit, or simplified azimuthal velocity, or axial 
through-flow profiles 
\cite{LinStab}, and later on for various aspect
ratios fixed by the diameters of both cylinders
\cite{CA-FluidMech-77,TJ-FluidMech-81,HM-Lond-77,DIPP-Lond-79}. 
Also, many experiments and comparisons
with the theoretical  predictions were done 
\cite{S-Lond-61,S-Ann-65,GuFahidy,GF-Can-86,TS3-93,%
TS-Euro-91,TS1-93,BAC-PhysRev-91,BCA-PhysD-92,BAC-Pre-93,%
TS-PhysRev-91,TGS2-93,Experiments,Kat-Jap-75}.

Our paper is organized as follows: In Sec.~\ref{system} we describe 
the system,
the subregions of absolute and convective instability, and our methods 
of investigation. Furthermore, we recapitulate the GLE. The next 
section presents the spatiotemporal behavior of PV patterns obtained 
numerically for two different boundary conditions. In 
Sec.~\ref{patternselection} 
we analytically and numerically elucidate the pattern selection 
observed within the GLE and the NSE. 
We compare the results with each other, with front propagation, and
with experiments. The last section gives a conclusion.

%%%%%%%%%%%%%%%%%%%%%%%%%%%%%%%%%%%%%%%%%%%%%%%%%%%%%%%%%%%%%%%%%%%%%%%
%
%  SECTION 2: The System
%
%%%%%%%%%%%%%%%%%%%%%%%%%%%%%%%%%%%%%%%%%%%%%%%%%%%%%%%%%%%%%%%%%%%%%%%

\section{The System}\label{system}

We investigate time-dependent, rotationally symmetric vortex structures 
in a Taylor-Couette apparatus with an externally enforced axial flow. 
The viscous, incompressible  fluid is
confined to the annulus between two concentric cylinders of inner radius
$r_1$ and outer radius $r_2$. The setup is characterized by two 
geometric
parameters: the radius ratio $\eta=\frac{r_1}{r_2}$ and the aspect ratio
$\Gamma$, i.e., the quotient of the axial extension of the annulus and 
the gap width $d=r_2-r_1$. Mostly we have used in our numerical 
simulations an
aspect ratio of $\Gamma=50$ and a radius ratio of $\eta=0.75$. The outer
cylinder is always kept at rest, while the inner one has a rotation rate
$\Omega_{cyl}$. 
In addition we impose a small through-flow in axial direction.
The boundary conditions at $r_1$ and $r_2$ were always no slip. The
conditions at the two ends $z=0$ and $z=\Gamma$ of the
annulus are explained in Sec.~\ref{boundaryconditions}. 

The flow pattern is described by the momentum balance equation for the
velocity field ${\bf u}$, the Navier-Stokes equations
\begin{mathletters} 
\begin{equation} 
\left( \partial_t + {\bf u}
\cdot{\bf  \mbox{\boldmath $\nabla$}} \right){\bf u} =
- \frac{1}{\varrho}{\bf \mbox{\boldmath $\nabla$}} p
+\nu{\bf \nabla}^2 {\bf u}\/, 
\end{equation}
and the continuity equation
\begin{equation}  
{\bf  \mbox{\boldmath $\nabla$}} \cdot {\bf u}=0
\end{equation}
\end{mathletters} 
which reflects the incompressibility of the fluid. Here $\nu$ is the
kinematic viscosity, $\varrho$ the mass density, and  $p$ the pressure.
The system is characterized by two dimensionless control parameters. 
The Taylor number
\begin{equation} 
T = \frac{\eta}{1-\eta} \frac{\Omega_{cyl}^2\,d^4}{\nu^2}
\end{equation} 
is given by the squared rotation rate $\Omega_{cyl}$ of the inner
cylinder. The Reynolds number 
\begin{equation} 
Re=\frac{\overline{w}\,d}{\nu}
\end{equation} 
is proportional to the mean axial through-flow velocity $\overline{w}$. 
For an axially uniform system the homogeneous basic flow state
\cite{C-Oxford-61}
\begin{equation} 
{\bf U} \left(r \right) = V_{CCF} \left(r \right)
{\bf e}_{\varphi} +  W_{APF} \left(r \right) {\bf e}_{z}
\label{BFLOW}
\end{equation} 
is a linear superposition of circular Couette flow (CCF)
\begin{equation}
V_{CCF} \left(r \right) = A r + \frac{B}{r} 
\end{equation} 
in azimuthal direction and of annular Poiseuille flow (APF)
\begin{equation} 
W_{APF} \left(r \right) = \frac{r^2 + C \ln r +  D}{E}Re \label{WAPF}
\end{equation}
in axial direction. We scale lenghts by $d$, times by the radial 
diffusion time $\frac{d^2}{\nu}$, azimuthal velocities by the velocity 
of the inner cylinder $\Omega_{cyl} r_1$, radial and axial velocities 
by $\frac{\nu}{d}$. Then 
\begin{mathletters} \begin{equation} 
A=-\frac{\eta}{1 + \eta}  \:, \quad B=-\frac{A}{\left(1-\eta\right)^2}
\end{equation}
\begin{equation} 
C= \frac{1 + \eta}{1 - \eta} \frac{1}{\ln \eta}  \:, \quad D= C
\ln(1-\eta) - \frac{1}{\left(1 -
\eta \right)^2}
\end{equation}
\begin{equation}  
E=-\frac{1}{2}\left[1+\frac{2 \eta}{\left(1- \eta \right)^2}+C \right].
\end{equation} \end{mathletters}
Here, $4 Re / E = \partial_z p$ is the dimensionless axial pressure
gradient driving the APF.

At the critical Taylor number $T_c \left(Re \right)$ 
\cite{RLM-PhysRevE-93,CA-FluidMech-77,TJ-FluidMech-81,BAC-PhysRev-91},
which depends on the through-flow rate, the basic flow becomes 
unstable to rotationally symmetric, axially extended PV perturbations 
via an oscillatory instability. There, a nonlinear PV solution
branches off the basic flow in an axially infinite system. We
consider the {\it deviation}
\begin{equation} 
{\bf u}
\left( r,z;t \right) = u {\bf e}_r + v  {\bf e}_{ \varphi} + 
w {\bf e}_z
\end{equation} 
of the velocity field from the basic flow (\ref{BFLOW}) as the
order-parameter field to characterize the secondary PV structure. 
We use the relative control parameter
\begin{mathletters}
\begin{equation}
\mu = \frac{T}{T_c \left(Re \right)} - 1  \quad
\end{equation}
corresponding to
\begin{equation}
\quad \epsilon= \frac{T}{T_c \left(Re=0 \right)} - 1
\end{equation}
\end{mathletters}
to measure the distance from the onset of PV flow for $Re \neq 0$ 
and of stationary Taylor vortex flow for $Re=0$, respectively. In 
this notation 
\begin{equation} 
\mu_c = 0 \quad \mbox{and} \quad \epsilon_c
\left(Re \right) =  \frac{T_c \left(Re \right)}{T_c \left(Re=0
\right)} -1
\end{equation}
is the critical threshold for onset of PV flow. The relation between 
$\mu$ and $\epsilon$ is 
\begin{equation}
\mu= \frac{\epsilon}{1+\epsilon_c\left( Re \right)}.
\end{equation}
The shear forces associated with the axial through-flow sligthly 
stabilize
the homogeneous basic state, so $\epsilon_c \left(Re \right)$
slightly increases with $Re$ 
\cite{RLM-PhysRevE-93,BAC-PhysRev-91,BCA-PhysD-92,BAC-Pre-93}.
For similar reasons a lateral Poiseuille shear flow suppresses the 
onset
of convection rolls with axes perpendicular to the flow in a B\'enard
setup of a fluid layer heated from below \cite{Kelly-94,MLK-Euro-89}.

\subsection{Ginzburg-Landau description}

Close to the bifurcation threshold $T_c \left(Re \right)$ of PV~flow,
i.e., for small $\mu$, the flow has the form of a harmonic wave, e.g.,
\begin{equation}
 w\left(r,z;t\right) = A\left(z,t\right) e^{i
\left(k_c z - \omega_c t\right)} \hat{w}\left(r\right) + c.c.
\label{HWAVE} 
\end{equation}
with a complex amplitude $A(z,t)$ that is slowly varying in $z$ and $t$.
The critical wave number $k_c$, frequency $\omega_c$ 
\cite{CA-FluidMech-77,TJ-FluidMech-81,RLM-PhysRevE-93,BAC-PhysRev-91,%
BCA-PhysD-92,BAC-Pre-93}, and eigenfunction
$\hat{w}(r)$ \cite{CA-FluidMech-77,AR-Dipl-91} appearing in 
Eq.~(\ref{HWAVE}) have been obtained from a
linear stability analysis of the basic flow state as functions of $Re$. 
The complex vortex amplitude
$A\left(z,t \right)$ is given by the solution of the 1D complex GLE
\begin{eqnarray} 
\tau_0\left(\dot{A}+v_g A'\right) & = & 
\mu\left(1+ic_0\right) A +  \xi^2_0\left(1+i c_1\right)A'' 
\nonumber \\
& & - \gamma\left(1+ic_2\right)\left |A\right|^2 A.  \label{GLE}
\end{eqnarray} 
Dot and primes denote temporal and spatial derivatives in the
$z$-coordinate, respectively. All coefficients of the GLE have been
calculated \cite{RLM-PhysRevE-93} as functions of  $Re$ for several 
radius ratios $\eta$.  As a consequence of the system's invariance 
under the combined symmetry operation $\left\{z \rightarrow -z,Re
\rightarrow -Re\right\} $ the coefficients $\tau_0$, $\xi^2_0$,
$\gamma$ are even in $Re$ while  the group velocity $v_g$ and the
imaginary parts $c_0$, $c_1$, $c_2$  are odd in $Re$ 
\cite{MLK-PhysRevA-92,RLM-PhysRevE-93}. 

It should be noted that the control parameter range of $\mu$ over which
(\ref{HWAVE}) gives an accurate description of the full velocity field 
of PV flow, say, on a percent level is indeed very small: The
asymmetry between radial in- and outflow intensities rapidly grows with
$\mu$ and causes higher axial Fourier contributions $\sim e^{inkz}$
\cite{HCAJ-PhysFluids-88,LR-ZPhysB-90,RLKS-PLENUM-92} 
to the velocity field that are
discarded in the $\mu \rightarrow 0$ asymptotics of the GLE
approximation (\ref{HWAVE}). However, the modulus of
the first Fourier mode of the vortex structures agrees for $Re = 0$ as
well as for $Re \not= 0$ quite well with the one predicted by the GLE 
-- cf.~Sec.~\ref{boundaryconditions}. 
On the other hand, the PV structure selected according to the
GLE differs from the one resulting from the full field equations --
cf.~Sec.~\ref{patternselection}. 

% section IIB

\subsection{Absolute and convective instability}
For small $\epsilon$ and $Re$ the control parameter plane is divided
into three stability regimes - cf.~Fig.~\ref{controlparameterplane}
- characterized by different growth behavior of {\it linear}
perturbations of the basic flow state.  
Below the critical threshold $\epsilon_c \left(Re \right)$ 
for onset of PV flow
(dashed line in  Fig.~\ref{controlparameterplane}) any 
perturbation, spatially localized as well as extended, decays. This 
is the parameter regime of absolute stability of the basic state.

Perturbations of the basic state can grow only for
$\epsilon > \epsilon_c(Re)$. However, in the presence of through-flow 
one has to distinguish \cite{BersBriggs}
between the spatiotemporal growth behavior of spatially localized
perturbations and of spatially extended ones. The latter having a 
form $\sim e^{ikz}$ can grow above $\epsilon_c$ (Re) -- in fact 
$\epsilon_c$ is determined as the stability boundary of the basic 
state against extended harmonic perturbations. On the other hand, 
a spatially localized
perturbation, i.e., a wave packet of plane wave perturbations is 
advected in the so-called convectively unstable parameter regime 
faster downstream than it grows -- while growing in the comoving 
frame it moves out of the system
\cite{BersBriggs,Huerre,Deissler,TES-PhysRev-90}. Thus, the 
downstream
as well as the upstream facing intensity front of the vortex
packet move into the same direction, namely, downstream. 
In this regime the flow pattern that results from a spatially
localized source, which generates perturbations for a limited time
only, is blown out of any system of finite length and the basic 
state is reestablished.

In the absolutely unstable regime (shaded region
in Fig.~\ref{controlparameterplane}) a localized
perturbation grows not only in downstream direction but it grows and
spatially expands also in upstream direction until the  upstream
propagating front encounters the inlet in a finite system. The final
pattern resulting in such a situation shows in downstream direction 
a characteristic axial intensity profile under which the PV flow 
develops with increasing distance from the inlet.

Within the framework of the
amplitude equation the boundary (full line in 
Fig.~\ref{controlparameterplane})  between absolute and convective
instability is given by \cite{Deissler}
\begin{equation} \mu^{c}_{conv} = \frac{\tau^2_0 v^2_g}{4 \xi^2_0
\left(1+c^2_1\right)}
\label{MUCCONV}
\end{equation}
corresponding to $\epsilon^{c}_{conv} =
\epsilon_c + (1+\epsilon_c) \mu^{c}_{conv}$. 
Thus the absolutely unstable regime 
(shaded region in Fig.~\ref{controlparameterplane}) is characterized by
$\mu > \mu^{c}_{conv}$ or, equivalently, by the reduced group velocity 
\begin{equation}
V_g = \frac{\tau_0}{\xi_0 \sqrt{\left(1+c^2_1\right)\mu}}
v_g = 2\sqrt{\frac{\mu^{c}_{conv}}{\mu}}
\label{Vg}
\end{equation}
being smaller than 2.

It should be mentioned that the GLE approximation (\ref{MUCCONV}) 
of $\mu^{c}_{conv}$ describes the boundary between absolute and
convective instability resulting from the NSE 
\cite{BAC-Pre-93,RDprivate}
very well for the small Reynolds numbers considered here.

% section IIC

\subsection{Methods of investigation}

The linear growth analysis of modes  
$exp\,[ i ( k z-\omega t + m \varphi)]$
shows that for small through-flow rates the homogeneous basic state 
becomes first unstable to axisymmetric PV flow
\cite{CA-FluidMech-77,TJ-FluidMech-81,NT-Proc-82}.
Up to $Re=4$ these patterns are also detected experimentally 
\cite{TS3-93,TS-Euro-91,TS1-93,BAC-PhysRev-91,BCA-PhysD-92,BAC-Pre-93}.
However, at larger $Re$ one observes stationary spirals and mixed 
flow patterns \cite{LDM-PhysFluids-92,TS3-93} in addition to 
the bifurcation of propagating spirals 
\cite{TS3-93,S-Lond-61,S-Ann-65,GuFahidy,GF-Can-86,Kat-Jap-75}.
The latter are predicted by the linear stability analysis to branch
off the basic state at $Re \approx 20$
\cite{CA-FluidMech-77,TJ-FluidMech-81,NT-Proc-82}.
The PV patterns occuring at small through-flow rates ($Re<4$) that 
are discussed here are rotationally symmetric. Therefore it is 
sufficient to solve the hydrodynamic field equations in an $r$-$z$ 
cross section of the annulus to describe the resulting field 
${\bf u}(r,z;t)$. 

We have performed numerical simulations of the 2D NSE. They are 
compared with analytical and numerical results obtained from the 1D 
GLE. The latter was solved with a Cranck-Nicholson algorithm using 
central differences for
spatial derivatives with a resolution of $20$ grid points 
per unit length $d$. The solution of the NSE was obtained with a 
time-dependent finite-differences marker and cell (MAC) algorithm
\cite{LMW-PhysRevA-85,Welch} 
with pressure and velocity being iteratively adapted to each 
other with the method of artificial compressibility 
\cite{Hirt}. Also here the spatial
resolution was $20$ grid points per unit length $d$. The temporal 
stepsize was $1/1800$ times the radial diffusion time $d^2/\nu$. 

When comparing finite-differences solutions of the NSE with 
experiments or with analytical properties, e.g., of the GLE we 
take into account that the
critical properties of the finite-differences MAC~code differ slightly 
from the latter due to its finite spatiotemporal resolution.
In particular the critical Taylor number, $T_c (Re)$, of the
MAC code lies slightly
(less than $1.5 \%$) below the theoretical bifurcation threshold
-- cf.~Table I and \cite{RLM-PhysRevE-93}. 
To find the marginal stability
curve $T_{stab} (k)$ of the MAC~algorithm for axially extended
PV~perturbations of wave number $k$ we analyzed the complex growth 
rate $s(k)$ in systems of length $2 \pi / k$ using periodic boundary 
conditions in axial direction for through-flow rates up to $Re=5$. 
From this analysis
we also obtained the critical values of the frequency $\omega_c$, group
velocity $v_g$, wave number $k_c$, and the parameters $\tau_0$ and
$\xi^2_0$ (Table I). Furthermore, we have cross-checked these results 
by investigating the evolution of localized perturbations in long 
systems of lengths $\Gamma > 50$, in
particular in the convectively unstable regime in a manner that 
is quite similar to experimental procedures
\cite{BAC-Pre-93,TS1-93}:
We generated tiny, localized PV perturbations of about five vortex pairs
with wave numbers close to $k_c$ under an intensity envelope of 
Gaussian shape.  They propagate downstream with the group velocity 
$v_g$. A fit to the rate of exponential growth of the envelope 
gives $\mu/\tau_0$. From the growth
of the full width at half maximum we obtain $\xi^2_0 /\tau_0$. These
identifications are based upon comparing with the temporal evolution 
of a Gaussian perturbation \cite{Deissler}
\begin{equation}
A(z,t=0) \propto exp[{-\frac{z^2}{2\Delta^2_0}}]
\end{equation}
of initial axial extension $\Delta_0$ according to the linearized GLE. 
It yields
\begin{equation}
A(z,t) \propto \frac{1}{\Delta (t)}
\,exp\,[\mu ( 1+i c_0) \frac{t}{\tau_0} -
\frac{(z-v_g t)^2}{2 \Delta^2 \left(t \right)}]
\end{equation}
with 
\begin{equation}
\Delta^2 \left( t \right)= \Delta^2_0 + 2 \xi^2_0
\left(1+i c_1 \right) \frac {t}{\tau_0}.
\end{equation}
Test runs with twice the lattice points
per unit length showed that the differences between the critical
properties of the
continuous system that were obtained \cite{RLM-PhysRevE-93} with a 
shooting method and 
the finite-differences system significantly decrease by one order 
of magnitude.

We have also investigated the dependence of nonlinear, saturated 
PV~flow on the numerical discretization. 
These tests show that the nonlinear vortex
structures are basically independent of the discretization -- provided 
it is not to coarse. 
However, one should base the comparison of bifurcated
flow structures obtained with different discretizations on the relative
control parameter $\mu = \frac{T}{T_c} - 1 $ that is influenced via 
$T_c$ by the discretization in question -- see also 
Sec.~\ref{finitesize}. 

When comparing results obtained for different $\mu$ and $Re$ from
numerical simulations of the NSE with those following from the GLE 
we found it sometimes advantageous to present them as functions of 
the reduced group velocity $V_g$ (\ref{Vg}). The deviation
\begin{equation}
2-V_g = 2 - 2 \sqrt{\mu^{c}_{conv} / \mu} 
\end{equation} 
is a scaled distance from the boundary between the absolutely and
convectively unstable regime. Its relation to the through-flow 
Reynolds number $Re$ and the Taylor number $T=(1+\mu) T_c(Re)$ can 
be read off from Table I. 

%%%%%%%%%%%%%%%%%%%%%%%%%%%%%%%%%%%%%%%%%%%%%%%%%%%%%%%%%%%%%%%%%%%%%%%
%
%  SECTION III: Propagating Vortex Patterns
%
%%%%%%%%%%%%%%%%%%%%%%%%%%%%%%%%%%%%%%%%%%%%%%%%%%%%%%%%%%%%%%%%%%%%%%%

\section{Propagating Vortex Patterns}\label{boundaryconditions}

We investigate PV~patterns in the absolutely unstable regime in an 
annulus of finite length with two different end conditions: a basic 
state boundary
condition (BCI) in Sec.~\ref{BSBC} and an Ekman vortex boundary 
condition
(BCII) in Sec.~\ref{EVBC}. For both boundary conditions numerical
simulations have been performed for the parameter combinations marked 
by symbols in Fig.~\ref{controlparameterplane}. 
In each case we observe a pattern of vortices
propagating downstream under a stationary intensity envelope after
transients have died out as shown in Fig.~\ref{hiddenlineplot}. The 
oscillation frequency $\omega$ of the flow is independent of the 
radial and axial position while the
local wave number $k$, the phase velocity $v_p=\omega/k$, and the 
vortex flow intensity varies with $r$ and $z$. The
frequency and the spatial variation of the PV pattern depends on the 
control parameters and on the boundary conditions but not on parameter 
history or initial conditions.

\subsection{Basic state boundary condition -- BCI}\label{BSBC}

Here we discuss vortex suppressing boundary conditions that are realized
by imposing the homogeneous basic state ${\bf U}(r)$ (\ref{BFLOW}) at 
in- and outlet of the annulus. The flow resulting for this boundary 
condition resembles the experimental one of Babcock 
{\em et al.}~\cite{BAC-Pre-93}. There, a system of flow distributors 
and meshes at 
the inlet reduces external perturbations penetrating the interior of 
the annulus and seems also to suppress radial flow \cite{BAC-Exp}. Near 
the inlet no {\it stationary} Ekman vortices are visible in this 
experiment and the amplitude of the downstream propagating vortices 
starts with zero or nearly zero 
\cite{BAC-PhysRev-91,BCA-PhysD-92,BAC-Pre-93}.

To illustrate the global properties of the flow patterns we present in 
the upper part of Fig.~\ref{hiddenlineplot} a hidden-line plot of the 
axial velocity field
$w$. Thin lines show snapshots of $w$ at the radial position 
$r=r_1+0.225$
obtained from the NSE at successive, equidistantly spaced times after
transients have died out. Then vortices propagate downstream under a
stationary intensity envelope (thick line). This envelope is determined
by the temporal extrema of $w(r_1+0.225;t)$ at any z-position.

The oscillation frequency of the PV~pattern is constant over the entire
annulus. Sufficiently away from inlet and outlet we observe a bulk 
region of nonlinear saturated PV flow with spatially uniform amplitude 
and wavelength. With increasing through-flow this bulk pattern is 
pushed 
further and further downstream. The growth length $l$ from the inlet 
over which the amplitude reaches half its saturation value depends on 
$\mu$ and $Re$.
It increases and finally diverges when the control parameters $\mu$, 
$Re$
approach the absolute-convective instability border $\mu^{conv}_c$, that
is when $V_g$ (\ref{Vg}) reaches $2$.

In Fig.~\ref{Lscaled} we compare the scaled growth length
\begin{equation}
L=\sqrt{\mu} \, l/\xi_0
\label{L}
\end{equation}
computed from the amplitude equation (solid line) with results from the
NSE (symbols) for various combinations of $\epsilon$ and $Re$. Due to 
the scaling property of the GLE 
\cite{MLK-Euro-89,MLK-PhysRevA-92,MLK-Nato-92} 
keeping in mind the smallness of the imaginary parts $c_i$ 
all values for $l$ obtained by the GLE subject to BCI fall onto 
one curve in the plot of $L$ vs.\ $V_g$. The open symbols representing 
the NSE results for BCI lie very close to the GLE curve. Full symbols 
for BCII are discussed in Sec.~\ref{EVBC}.

For further characterization of the PV~flow structure we found a
{\it temporal} Fourier decomposition of the time-periodic fields, e.g.,
\begin{equation}
w(r,z;t)= \sum_{n} w_{n} \left( r,z \right) e^{-i n \omega t}
\end{equation}
to be useful. To that end we first determined the oscillation frequency
$\omega$ of the PV~pattern and checked that it showed no spatial 
variation.  In Fig.~\ref{fouriermodesBSBC}a the zeroth temporal Fourier 
mode $w_0$ and the moduli $|w_1|$, $|w_2|$, and $|w_3|$ resulting from 
the NSE are shown by 
full lines while $|w_1|$ from the GLE is shown by the dashed one. The 
latter was rescaled by a factor of $\approx 0.94$ which comes from 
comparing 
the first axial Fourier mode of $w(r=r_1+0.225)$ at $Re=0$ in an axially
periodic system with the
GLE result of Recktenwald 
{\it et al.}~\cite{RLM-PhysRevE-93,AR-Dipl-91}. 
As an aside we mention that for the radial
velocity $u$ having a nodeless radial profile with a single maximum 
the deviation of the first 
Fourier mode of $u(r=r_1+0.5)$ between finite-differences NSE and GLE 
is only $0.8\%$ at $Re=0$.
 
Note that the
solution of the GLE (\ref{GLE}) 
\begin{equation} \label{HWAVENOTRANSIENTS}
A(z,t)= R(z) e^{i [ \varphi \left( z \right) - \Omega t ]}
\end{equation}
oscillates harmonically with frequency $\Omega$ under a stationary
envelope 
\begin{equation}
R(z)= |A(z,t)|
\end{equation}
after transients have died out. Therefore, the GLE velocity field
(\ref{HWAVE}) contains no temporal Fourier mode other than $n=1$ whereas
the solution of the full NSE has higher harmonics.

In the bulk region the {\it temporal} modes obtained for finite 
through-flow from the NSE grow for small $\mu$ proportional to 
$\mu^{n/2}$ with relative
corrections proportional to $\mu$. Thus, they show the same growth
behavior with $\mu$ that the {\it axial} Fourier modes 
\cite{HCAJ-PhysFluids-88,D-FluidMech-62,KRL-PhysRevA-89,LR-ZPhysB-90,%
RLKS-PLENUM-92} of
stationary Taylor vortices 
without through-flow show as a function of $\epsilon$. The reason is 
that all fields in the PV~state have the form of propagating waves
\begin{equation}
f_b(r,z;t)= f_b (r,z-\frac{\omega}{k_b} t )
\end{equation}
in the bulk region - denoted by a subscript b. There the wave number 
$k_b$ and the phase velocity $\omega / k_b$ do not vary with $z$. 

The Fourier mode $w_0$ of the PV~pattern is not zero everywhere
since the variation of the flow amplitude under the fronts causes a
secondary, stationary, closed flow pattern there
-- see, e.g., $w_0$ at $z \approx 9$ and near the outlet in 
Fig.~\ref{fouriermodesBSBC}a. Its strength and
structure depends on the steepness of the front. For example, an 
increase
of $Re$ leads to a smaller gradient of the upstream facing front and to
weaker stationary secondary flow extending further downstream. 

The intensity variation 
(cf.~Figs.~\ref{hiddenlineplot}, \ref{fouriermodesBSBC}a) of
PV flow under the fronts also causes there a spatial variation in the 
local
wavelength $\lambda (z)$ (Fig.~\ref{fouriermodesBSBC}b) and in the 
phase velocity $v_p(z)= \omega \lambda (z) / 2\pi$ with the 
oscillation
frequency $\omega$ of PV~flow being constant. The full (dashed) line 
in Fig.~\ref{fouriermodesBSBC}b is the wavelength profile 
$\lambda (z)$
selected for BCI according to the NSE~(GLE). This figure shows that 
the GLE does not describe the pattern selection quantitatively. 
The selected wavelength in the bulk region of the NSE~(GLE) is 
smaller ~(larger) than the critical one ~-- see also 
Sec.~\ref{patternselection}.

The local wavelength shown in Fig.~\ref{fouriermodesBSBC}b was 
determined
from the phase gradient of the first temporal Fourier mode $w_{1} (z)$ 
at the radial position $r=r_{1}+0.225$. We have analyzed $w_1$ and 
also $u_1$ at other $r$ positions and in addition we determined the 
axial distances between node positions of $u$ and $w$. Under the 
fronts there is a significant
radial variation of $\lambda$. Furthermore, the wavelengths determined 
via the phase gradients of $w_1$ and $u_1$ differ there from each
other and from those obtained via node distances of $w$ and $u$. But in
the bulk part of the PV~structure all these quantities yield the same
wavelength $\lambda_b$.

% section IIIB

\subsection{ Ekman vortex boundary condition -- BCII \label{EVBC}}

In Taylor-Couette experiments without through-flow rigid nonrotating
end plates bounding the fluid in axial direction induce stationary
Ekman vortex flow 
\cite{ACDH-PhysD-86,HAC-PhysRev-86,PR-PhysLettA-81}.
These vortices are also detected by Tsameret {\em et al.}
in experiments with axial through-flow, where meshes at
the apertures were used as nonrotating boundaries at in- and outlet
\cite{TS-Euro-91,TS1-93}.

The second axial boundary condition BCII 
enforces stationary Ekman vortices near the apertures of
the annulus. To that end we impose at both ends $z=0, \/ \Gamma$
zero radial flow, zero azimuthal flow, and in axial direction the
annular Poiseuille flow $W_{APF}(r)$ (\ref{WAPF}). The
spatiotemporal properties of the vortex pattern subject to this
BCII can be seen in the lower part of Fig.~\ref{hiddenlineplot}. 
There we show
a hidden-line plot of the axial velocity in the same way and for
the same parameters as for the BCI pattern in the upper part of
Fig.~\ref{hiddenlineplot}.

In the immediate vicinity of in- and outlet there
are {\it stationary}~Ekman vortices whose intensity rapidly
decreases towards the bulk of the annulus. In addition there is
--~as in the BCI case~-- the PV flow structure that
{\it oscillates}~in time with a z-independent frequency $\omega$.
The oscillation amplitude drops to zero near the apertures due to the 
boundary conditions, and the stationary Ekman vortices increase the 
growth length $l$ of the oscillating structure in comparison to 
the BCI case. 

These 
two different flow elements are best separated by a temporal Fourier
analysis. The results for BCII are shown in 
Fig.~\ref{zerothfouriermodeEVBC} for different through-flow rates. 
The zeroth temporal mode $w_0$ (thick line)
represents the stationary Ekman vortex flow of the system. The
temporally oscillating PV~structure is characterized by the modes
$w_n$ (thin lines). The Ekman vortex structure at the inlet is only
slightly affected by the through-flow: the strength of the Ekman
vortex closest to the inlet decreases somewhat with increasing $Re$, 
while its extension slightly increases. The stationary vortices
at the outlet become more and more squeezed together with increasing
$Re$, and also their intensity reduces. On the other hand, the
oscillatory pattern of PV responds dramatically to the through-flow
in being more and more pushed downstream. Note also that for small
$Re$ where the PV~amplitude of, say, $\vert w_1 \vert$ overlaps near
the inlet with the Ekman vortex intensity the latter causes axial
oscillations in the harmonics of the PV~flow and similarly near the
outlet - cf.~Fig.~\ref{zerothfouriermodeEVBC}.

The filled symbols in Fig.~\ref{Lscaled} represent the growth
length of the PV~flow intensity. This length diverges in a similar 
way at the convective instability border $V_g = 2$ as the one obtained 
for BCI. But due to the presence of Ekman vortices at the inlet the
characteristic growth length  $l(\epsilon,Re)$ of PV flow is increased
by $0.5-2.5 d$ in comparison to the BCI case -- Ekman vortices
push the PV~structure further downstream. The difference
$l_{BC II} - l_{BC I}$ decreases with increasing through-flow rate
as Ekman vortices and PV~flow become more and more separated from
each other.

As in the BCI case the zeroth temporal Fourier mode exhibits a 
secondary, stationary, closed flow pattern of low intensity under 
the fronts of the PV~pattern. For small 
Reynolds numbers this pattern is concealed by the Ekman vortex flow, 
whereas it becomes visible for bigger through-flow rates.

%%%%%%%%%%%%%%%%%%%%%%%%%%%%%%%%%%%%%%%%%%%%%%%%%%%%%%%%%%%%%%%%%%%%%%%
%
%  SECTION IV: Pattern selection
%
%%%%%%%%%%%%%%%%%%%%%%%%%%%%%%%%%%%%%%%%%%%%%%%%%%%%%%%%%%%%%%%%%%%%%%%

\section{Pattern selection}\label{patternselection}

Our investigations of the PV structures show that the axial
through-flow causes a unique pattern selection such that the selected
structure is independent of history and initial conditions and depends
only on the control parameters and boundary conditions. Thus, the axial
flow causes possible wave numbers within the Eckhaus-stable band of 
stationary
Taylor vortex patterns without through-flow to collapse to only one
uniquely selected PV structure for nonvanishing through-flow rates. 
However, the PV patterns that
are selected according to the GLE approximation and to the full NSE 
differ
distinctly from each other. We first investigate in
Sec.~\ref{patternselectiona} the selection mechanism within the GLE 
framework. In Sec.~\ref{patternselectionb} we compare with NSE results 
and experiments. 

\subsection{Pattern selection within the GLE}\label{patternselectiona}

Here we elucidate how the pattern selection mechanism of the GLE can 
be understood as a nonlinear eigenvalue/boundary-value problem where 
the frequency of the PV pattern is the eigenvalue. Thus, the selected
PV~structure is characterized and determined by the combination of
eigenvalue and corresponding eigenfunction.

\subsubsection{The eigenvalue problem}

We look for solutions of the GLE (\ref{GLE}) 
of the form (\ref{HWAVENOTRANSIENTS})
\begin{mathletters}
\begin{equation} 
A(z,t) = a(z)e^{-i\Omega t} =  R(z) e^{i[\varphi (z) - \Omega t]}
\label{EVPROBLEMA}
\end{equation}
with stationary envelope $R(z)$, stationary wave number
\begin{equation} 
q(z) = k(z)-k_c = \varphi'(z),\label{EVPROBLEMB}
\end{equation}
and constant frequency
\begin{equation}
\Omega = \omega - \omega_c. \label{EVPROBLEMC}
\end{equation}
Inserting this solution ansatz 
into the GLE (\ref{GLE}) one obtains the nonlinear eigenvalue 
problem
\begin{eqnarray}
\tau_0\left(-i\Omega a+v_g a'\right) & = & 
\mu\left(1+ic_0\right) a +  \xi^2_0\left(1+i c_1\right)a''
\nonumber \\
& & - \gamma\left(1+ic_2\right)\left |a\right|^2 a  
\label{EVPROBLEMD}
\end{eqnarray}
or, equivalently,
\begin{eqnarray}
i\tau_0(-\Omega+v_g q)R+\tau_0 v_g R'  =  \mu(1+ic_0)R  \nonumber\\
+\xi^2_0(1+ic_1) (R''-q^2R+iq'R+2iqR') -\gamma(1+ic_2)R^3 \nonumber \\
\label{EVPROBLEME}
\end{eqnarray}
\label{EVPROBLEM}
\end{mathletters}
as a solvability condition with $\Omega$ being the eigenvalue. We are
interested in solutions $R(z)$ and $q(z)$ that look like the  dashed 
lines in Fig.~\ref{fouriermodesBSBC}, i.e., where the variation of 
$R(z)$ and in 
particular of $q(z)$ is "as small as possible". 
Such a solution type seems to be connected with the
eigenvalue $\Omega$ that is closest to zero. Here we consider
the boundary conditions $R_{in,out}=0$ at in- and outlet. These
requirements fix four boundary conditions
-- namely, $Re(A_{in,out}) = Im(A_{in,out}) =0$ -- which are necessary
to solve the GLE (\ref{GLE}).

Note that the eigenfrequency $\Omega$ is in general determined by
{\em global} properties and not by the {\em local} variation of the 
eigenfunctions $R(z)$ and $q(z)$, say, at the inlet or 
in the bulk. Nevertheless it is informative and useful for our further
discussion to list below relations between $\Omega$ and structural
properties at the inlet and in the bulk of the annulus.

\subsubsection{Inlet behavior}

From (\ref{EVPROBLEME}) one finds that the
BCI at in- and outlet, $R=0$, implies the relation
\begin{equation}
R''_{in,out} =[\frac{\tau_0 v_g}{\xi^2_0(1+ic_1)}
-2iq_{in,out}]R'_{in,out}
\label{R2R1COMP}
\end{equation}
at $z=0,\Gamma$. Hence, whenever the modulus grows with finite slope,
$R'_{in,out}\neq0$, the boundary condition $R=0$ fixes the wave number 
at inlet {\em and \/} outlet to the value $q_{in}=q_{out}$
\begin{equation}
\xi_0 q_{in,out}=-\frac{1}{2} c_1\frac{\tau_0v_g}{\xi_0 (1+c^2_1)}
=-\frac{c_1}{\sqrt{1+c^2_1}}\sqrt{\mu^c_{conv}} 
\label{QINOUT} 
\end{equation}
that depends only on $Re$. This value ensures that the imaginary part of
the square bracket in (\ref{R2R1COMP}) is zero. 
For $Re=0$ one obtains $q_{in,out}=0$ as Cross 
{\em et al.}~\cite{CDHS-FluidMech-83}.
In addition the condition 
$R=0$ yields via the real part of (\ref{R2R1COMP}) to two relations
\begin{equation}
R''_{in,out}=\frac{\tau_0v_g}{\xi^2_0 (1+c^2_1)}R'_{in,out}
\label{R2R1}
\end{equation}
between the different slopes $R'_{in} \neq R'_{out}$ and curvatures
$R''_{in} \neq R''_{out}$ of the modulus at the two respective 
boundaries that hold independently of $\mu$.

Furthermore, one can derive a relation between the eigenfrequency
$\Omega$, the wave number $q_{in,out}$, and its slope  $q'_{in,out}$ 
at inlet and outlet. To that end we consider the axial derivative of
(\ref{EVPROBLEME}) at inlet and outlet with $R=0$
\begin{eqnarray}
i\tau_0(\Omega - v_g q) + \mu (1+ic_0) - \xi^2_0 (1+ic_1)q^2
\nonumber \\
-[\tau_0 v_g -2iq \xi^2_0 (1+ic_1) ] \frac{R''}{R'}
=\xi^2_0 (1+ic_1) (\frac{R'''}{R''}+3iq'). \nonumber \\
\end{eqnarray}
This allows to solve for $\frac{R'''}{R''}$ and $q'$ separately. 
Using (\ref{QINOUT},\ref{R2R1}) one obtains the relation
\begin{eqnarray}
\tau_0\Omega & = & (c_1-c_0)\mu + \tau_0 v_g q_{in,out} \nonumber \\
& & -3(1+c^2_1)\xi^2_0 q'_{in,out}. 
\label{OMQINOUT}
\end{eqnarray}
Note that this equation implies $q'_{in}=q'_{out}$.

The relation (\ref{OMQINOUT}) also demonstrates the nonlocality of the
pattern selection mechanism and of the eigenvalue problem. At first 
sight one might be tempted to infer from the above relation between 
$\Omega$ and
$q_{in}, q'_{in}$ that the latter two fix the frequency. However,
Eq.~(\ref{OMQINOUT}) equally well holds for $q_{out},q'_{out}$ and 
the outlet properties do not fix $\Omega$ neither. So the correct 
interpretation of Eq.~(\ref{OMQINOUT}) is that the eigenvalue $\Omega$ 
being the characteristic
global signature of the pattern fixes the local quantities 
$q'_{in,out}$ for the given boundary conditions.

\subsubsection{Bulk behavior}

The wave number $q(z)$ away from in- and outlet in general differs 
from $q_{in}(=q_{out}$). In order to obtain the full axial profiles 
of the eigenfunctions $R(z)$ and $q(z)$ belonging to the eigenvalue 
$\Omega$ one
has to solve the eigenvalue problem (\ref{EVPROBLEM})
with the boundary condition
$R_{in,out}=0$. Alternatively one can solve numerically the 
time-dependent GLE (\ref{GLE}).

However, when the system size and the control parameters are such that 
the PV pattern forms a bulk part with a homogeneous modulus $R_b$ and 
wave number $q_b$, i.e., where 
\begin{mathletters} 
\begin{equation}
R'_b = R''_b = q'_b = 0.
\end{equation}
Then, the dispersion relation
\begin{equation}
\tau_0\Omega= (c_2-c_0)\mu + \tau_0 v_g q_b +  (c_1-c_2)\xi^2_0 q^2_b 
\label{OMQBULK}
\end{equation}
provides a relation between the frequency eigenvalue $\Omega$, the 
bulk wave number $q_b$, and the bulk modulus  
\begin{equation}
R^2_b=\frac{\mu-\xi^2_0 q^2_b}{\gamma}. \label{RQBULK}
\end{equation}
\end{mathletters}
The equation (\ref{OMQBULK}) establishes together with (\ref{OMQINOUT}) 
also a relation between $q_b$ on the one hand and $q_{in}=q_{out}$ and 
$q'_{in}=q'_{out}$ on the other hand. 

\subsubsection{Pattern selection at the convective instability boundary}

Approaching the convective instability boundary (\ref{MUCCONV}) we 
found that $q'_{in}$ decreases to zero so that the frequency 
eigenvalue becomes
\begin{equation}
\tau_0\Omega^c_{conv}=- (c_0+c_1)\mu^c_{conv}. \label{OMCCONV}
\end{equation}
Then the bulk wave number approaches according to (\ref{OMCCONV}, 
\ref{OMQBULK}) the limiting value
\begin{equation}
\xi_0 (q_b)^c_{conv}=
-\frac{\sqrt{1+c^2_1} {(+)\atop-} \sqrt{1+c^2_2} }{c_1-c_2}
\sqrt{\mu^c_{conv}},
 \label{QCCONV}
\end{equation}
while the inlet wave number 
$q_{in}$ is given by (\ref{QINOUT}). Due to the very small imaginary 
parts $c_i$ the solution (\ref{QCCONV}) with the plus sign has to be 
discarded, since it corresponds to unphysically large wave numbers.

In Fig.~\ref{selectedwavelength} the bulk wavelengths selected 
according to the GLE are
shown by halftone symbols. The results in the absolutely unstable 
regime,
$V_g<2$, were obtained numerically by integrating the time-dependent
GLE (\ref{GLE}). The value shown right at the border,
 $V_g=2$, of the absolutely unstable regime is the analytical expression
(\ref{QCCONV}). Note that the selected GLE wavelengths are mostly 
larger than the critical one.

\subsubsection{Comparison with front propagation}

It is suggestive to compare properties of PV~patterns like the one in 
the upper part of Fig.~\ref{hiddenlineplot} with those behind an
upstream facing front 
that connects in an axially infinite system the basic state at 
$z = - \infty$ with the developed
PV~state at $z = + \infty$. A {\it linear} growth analysis of the GLE
(\ref{GLE}) along the lines of \cite[Sec.~VI B 3]{CH-RevModPhys-93} 
shows that the far tail of a {\it linear} front of PV perturbations
moves with the front velocity
\begin{equation}
v_F = v_g - 2 \frac{\xi_0}{\tau_0} \sqrt{\mu (1+c_1^2)} .
\end{equation}
In the convectively unstable regime $0 < \mu < \mu_{conv}^c$ the front
propagates downstream so that $v_F > 0$. In the absolutely unstable 
regime $\mu > \mu_{conv}^c$ it moves upstream, i.e., $v_F < 0$. And 
at the boundary
$\mu = \mu_{conv}^c$ the front is stationary -- $ v_F = 0$. 
We compare the
properties of such a stationary front at $\mu=\mu_{conv}^c$ with those 
of the solution $A(z,t)$ (\ref{HWAVENOTRANSIENTS}) with stationary 
modulus in a semi-infinite system ($0 \leq z < \infty$). 
The reason for restricting the comparison to
$\mu_{conv}^c$ is that an upstream moving front in the absolutely 
unstable regime will be pushed against the inlet whence the 
spatiotemporal
structure of the free front gets modified by the inlet. 

The far tail of the stationary front resulting from the linear GLE 
is characterized by the local wave number
\begin{equation}
\xi_0 q_F (v_F = 0) = - \frac{c_1}{\sqrt{1+c^2_1}}\sqrt{\mu^c_{conv}} 
\label{QFLIN}
\end{equation}
and the oscillation frequency
\begin{equation}
\tau_0 \Omega_F (v_F = 0) = - (c_0 + c_1) \mu_{conv}^c. \label{OMFLIN}
\end{equation}
In fact in the linear part of the stationary front the flow amplitude
\begin{equation}
A(z,t) \sim e^{i [Q_s z- \zeta (Q_s) t]}
\end{equation}
varies with a wave number $Re(Q_s)= q_F$ (\ref{QFLIN}), a spatial decay 
rate $Im(Q_s)$, a temporal growth rate $Im(\zeta (Q_s))= 0$, and an 
oscillation frequency $Re(\zeta (Q_s)) = \Omega_F$ (\ref{OMFLIN}). Here
\begin{equation}
\xi_0 Q_s = - \frac{1}{2} \frac{i}{1+i c_1} \frac{\tau_0}{\xi_0} v_g
          =-\frac{i+c_1}{\sqrt{1+c^2_1}}\sqrt{\mu^c_{conv}} 
	  \label{QCOMPLEXFRONT}
\end{equation}
is the saddle position of the complex dispersion relation
\begin{equation}
\tau_0 \zeta (Q, \mu) = \tau_0 v_g Q + i(1+i c_0) \mu - i(1+i c_1) 
 \xi_0^2 Q^2 \label{DISPERSIONFRONT}
\end{equation}
of the linear GLE in the complex $Q$-plane determined by the 
condition
\begin{equation}
\begin{array}{rll}
{\displaystyle \frac{d \zeta (Q)}{d Q}  \bigg\vert} & & = 0  \\
 & \hspace{-0.15cm} $\raisebox{2.0ex}[-2.0ex]{$Q_s$}$  & \\
\end{array} 
\end{equation}
for $v_F = 0$, i.e., at $\mu = \mu_{conv}^c$.

So the inlet wave number $q_{in}$ (\ref{QINOUT}) agrees with $q_F$ 
(\ref{QFLIN})
and $\Omega_{conv}^c$ (\ref{OMCCONV}) coincides with $\Omega_F$ 
(\ref{OMFLIN}).
Thus, right at the border $\mu_{conv}^c$ of
the absolutely unstable regime the frequency eigenvalue of the PV 
pattern
growing in downstream direction from the inlet at $z=0$ agrees with the
oscillation frequency selected by the stationary, {\it linear} front. 
Within the front propagation point of view the PV pattern develops 
from the basic state with an exponential growth rate $\kappa$ of the 
modulus. This leads for $R\rightarrow 0$ to
$R'=\kappa R,R''=\kappa^2 R,...\rightarrow 0$ for 
$z \rightarrow -\infty$. On the other hand, the PV flow intensity in 
the 
semi-infinite system in general drops to zero at the inlet with finite 
$R'$ and $R''$. Only at the border line $\mu = \mu_{conv}^c$ the 
difference disappears since there $R',R'',...\rightarrow 0$ for $z=0$.  
In the absolutely unstable regime, $V_g < 2$, the frequency eigenvalues
$\Omega = \omega - \omega_c$ of the GLE lie slightly above the limiting 
value $\Omega_{conv}^c$ at the border of the absolutely unstable 
regime. 
The shaded symbols in Fig.~\ref{omvalues} show for three different 
$\epsilon$ how
$\Omega$ obtained numerically by integrating Eq.~(\ref{GLE})
approaches with increasing $Re$ the front frequency
$Re(\zeta (Q_s)) = \Omega_F (v_F = 0) = \Omega_{conv}^c$ (dashed line). 
For convenience, the merging points with this limit line are marked 
for the three $\epsilon$-values investigated here by small halftone 
symbols.

Using this frequency for the full nonlinear PV pattern at 
$\mu_{conv}^c$ one obtains from the bulk dispersion relation 
(\ref{OMQBULK}) an expression for the bulk wave number 
\begin{eqnarray}
\xi_0 (q_b)_F &  = & \xi_0 (q_b)_{conv}^c \nonumber \\ 
&  = & -\frac{\sqrt{1+c^2_1} {(+)\atop-} \sqrt{1+c^2_2} }{c_1-c_2}
\sqrt{\mu^c_{conv}} \label{QBULKFRONT}
\end{eqnarray}
far behind the front.

This bulk wave number behind the front has the opposite sign of the 
one that is cited in 
\cite{TS-Euro-91} and \cite[ Eqs.~(10,14)]{TS1-93}.
An explanation for this might be that the formulas
of Nozaki and Bekki \cite[Eqs.~2, 8, and 9]{NB-PhysRev-83}
that have been used in 
refs.~\cite{TS-Euro-91,TS1-93} 
do not seem to have been transformed to a 
through-flow situation with an {\it upstream} facing front that joins 
in upstream direction to the basic state, in our case at $z = - \infty$.
After correcting the sign error, the rough qualitative agreement 
that was
reported in \cite{TS-Euro-91,TS1-93} between the experimental wave 
numbers \cite{TS-Euro-91,TS1-93} and the erroneous GLE results 
\cite{TS-Euro-91,TS1-93} disappears.
In fact, both, the experiments \cite{TS-Euro-91,TS1-93} and our
numerical simulations of the NSE yield wave numbers that differ in a 
common distinctive way from the proper GLE result (\ref{QBULKFRONT}).

As an aside we mention, that the {\it nonlinear} front 
solution of Nozaki and Bekki \cite[Eqs.~2-7]{NB-PhysRev-83} 
yields for our GLE (\ref{GLE})
a front  that is not stationary at $\mu_{conv}^c$ but rather moves 
with velocity
\begin{equation}
v_{NB} = - ( 3 \sqrt{\frac{1+c_1^2}{8+9 c_1^2}} - 1 ) v_g
\end{equation}
in upstream direction with a bulk wave number
\begin{eqnarray}
\xi_0 (q_b)_{NB} = 
\frac{\sqrt{\mu^c_{conv}}}{(c_2 - c_1) \sqrt{8+9 c_1^2}} \nonumber \\
\times \Bigl[ 3 (1+c_1^2) + sign (c_2 - c_1) 
\sqrt{8  (c_2 - c_1)^2 + 9(1+ c_1 c_2)^2} \Bigr]. \nonumber \\
\end{eqnarray}
So we conclude that the {\it nonlinear} front solution of Nozaki and 
Bekki is
unrelated to the PV patterns at $\mu_{conv}^c$. Furthermore, it was 
noted \cite{SH-PhysD-92} that localized initial perturbations did
not evolve into the {\it nonlinear} front solution of Nozaki and Bekki
\cite{NB-PhysRev-83,NB-Jap-84,CM-PhysD-93}.
However, $v_{NB}$ and
$(q_b)_{NB}$ differ for small through-flow only slightly
from the respective values $v_F = 0$ and $(q_b)_F$.
For example at ~$Re=2$ one obtains $v_{NB}=-0.06 v_g$ and
$(q_b)_{NB}=1.15 (q_b)_F$.

\subsubsection{Scaling properties}

Scaling length and times according to
\begin{equation}
\hat z =  \sqrt{\mu}\frac{z}{\xi_0} \,\,;\,\, \hat t = \mu 
\frac{t}{\tau_0},
\end{equation}
the GLE for the reduced amplitude $\hat A=A\sqrt{\gamma/\mu}$ does no
longer contain $\gamma$ and $\mu$ explicitly but the reduced 
group velocity $V_g$ (\ref{Vg}) and the coefficients $c_0, c_1, c_2$. 
Thus the reduced selected frequency
$\hat \Omega=\tau_0\Omega /\mu$ and the selected bulk wave number
$\hat q_b =\xi_0 q_b /\sqrt{\mu}$ depends not only on $V_g$ but via
$c_0, c_1, c_2$ also on the Reynolds number. This $Re$-dependence can 
alternatively be seen -- via the $Re$-dependence of $V_g$ and of the
$c_i$'s entering
(\ref{Vg}) -- also as an additional $\mu$-dependence. 
The latter dependence is sufficiently strong to prevent a scaling 
of $\hat \Omega$
with $V_g$ alone. However, by reducing the eigenfrequencies
$\Omega (\epsilon, Re)$ selected for our three different
$\epsilon$-values by the limiting values
$\Omega (\epsilon, Re^c_{conv}) = \Omega^c_{conv}$ (\ref{OMCCONV}) at 
the border of the absolutely unstable regime we effectively have 
eliminated 
the $\epsilon$-dependence and all GLE~data (halftone symbols in 
Fig.~\ref{scaledom}) fall onto one curve. A similar scaling holds for 
the bulk wave number $q_b$ divided by $(q_b)^c_{conv}$ (\ref{QCCONV}).

Remarkably enough, by reducing the NSE frequencies with the GLE 
frequency
$\Omega^c_{conv}$ (\ref{OMCCONV}) the NSE results (open symbols in
Fig.~\ref{scaledom}) almost show this one-variable scaling with $V_g$.
However, for the GLE $(\omega - \omega_c)/\Omega^c_{conv}$ increases 
monotonously with the scaled
group velocity $V_g$ whereas the NSE results show a monotonous decrease 
with $V_g$. This discrepancy between  NSE and GLE results is caused by
the different dispersion relations of the full hydrodynamic equations 
and the approximate GLE -- cf.~below.

% IVB

\subsection{Pattern selection within the NSE} \label{patternselectionb}

Here we compare our numerical solutions of the NSE with results from the
GLE and from experiments. First we discuss the selected wave numbers in
the bulk region of PV~flow. Then in comparison with front propagation
we show common selection properties of NSE and GLE. Finally we compare 
with experiments.

\subsubsection{Bulk wavelengths -- NSE $vs$ GLE}

In Fig.~\ref{selectedwavelength} we show all selected bulk wavelengths 
obtained from numerical simulations of the NSE for BCI (open symbols) 
and BCII (filled symbols) in comparison with results obtained 
numerically from the GLE for BCI (halftone symbols). The NSE 
wavelengths that are selected
when the basic state is enforced at in- and outlet weakly decrease with
increasing $Re$ and $\epsilon$. For the smallest $\epsilon=0.0288$ their
values are very close to the critical wavelength $\lambda_c$ (dotted
line). For this boundary condition the wavelengths are mostly smaller 
than the critical one with a deviation of up to $2\%$ for the largest
$\epsilon = 0.186$ that was investigated here. These deviations are 
much stronger than the decrease of the critical wavelength with 
increasing Reynolds numbers. 

It is interesting to note that transversal Rayleigh-B\'enard convection
rolls in horizontal shear flow \cite{MLK-Euro-89,MLK-PhysRevA-92} show 
a qualitative
and quantitative \cite[$\epsilon = 0.114$]{MLK-PhysRevA-92} 
correspondence to PV~flow. Therefore, our results are not limited 
to PV~flow in a Taylor-Couette apparatus only. In this context 
we also mention the experimental
investigation of Gu and Fahidy \cite{GuFahidy}. 
They observed that
axial through-flow causes a rearrangement of the vortex centers similar 
to the one seen for tranverse convection rolls subject to a 
horizontal flow
\cite[Fig.~2]{MLK-PhysRevA-92}. Without through-flow the vortices are
aligned along a straight line roughly in the gap center. With 
increasing through-flow vortices are alternatingly displaced towards 
the outer or inner cylinder, thereby reducing the axial component 
of their distance and thus the wavelength. Viewed in the 
$z-r$~plane with $r_2$ above $r_1$,
right (left) turning vortices -- for which $w$ is negative (positive) 
close to the inner cylinder -- are displaced towards the inner (outer)
cylinder by a through-flow directed from left to right, since the
through-flow enhances the axial flow component $w$ of a right turning 
vortex near the outer cylinder at $r_2$, and weakens it near the 
inner one at $r_1$ and vice versa for the left turning one.

Obviously the wavelengths resulting for 
BCI from the NSE (open symbols in Fig.~\ref{selectedwavelength}) differ 
substantially and systematically from those
resulting from the GLE (halftone symbols): the former decrease with 
growing $Re$ and
$\epsilon$ while the latter increase. Nevertheless we think --
cf.~the discussion in Sec.~\ref{comparisonfront} -- that the 
selection mechanism is quite similar.

\subsubsection{The effect of Ekman vortices}

The NSE wavelengths obtained for the Ekman vortex generating BCII 
(filled symbols in Fig.~\ref{selectedwavelength}) are similar 
to the BCI results in the limiting cases of small $Re$ or 
small $\epsilon$, and near the border line
$V_g = 2$ of the absolutely unstable regime. In the vicinity of 
$V_g = 2$ the stationary Ekman vortices and the PV flow are spatially 
separated, and their interaction is very weak allowing for PV 
wavelengths like those for BCI. For very small Reynolds numbers 
the agreement in the PV wavelengths
for BCI and BCII is not yet understood. In between there is a visible
difference of up to $5 \%$ ($\epsilon=0.186$ and $V_g \approx 0.4$)
between the bulk wavelengths obtained for the two boundary conditions.

However, we found that the bulk phase velocities, $v_p = \omega / k$, 
were almost independent of the boundary conditions: the relative 
difference of
$v_p$ for BCI and BCII is less than $0.1 \%$ for the parameter regimes
investigated here. Furthermore, the deviation of $v_p$ from the 
critical value $(v_p)_c = \omega_c/k_c$ is less than about $1 \%$. 
Hence, one can 
infer from the plots in Fig.~\ref{selectedwavelength} for
$\lambda / \lambda_c - 1$ also the variation of the frequency 
eigenvalue with through-flow since
\begin{equation}
\frac{\omega}{\omega_c} - 1 = \frac{v_p}{(v_p)_c}
\frac{k}{k_c} - 1 \, \simeq \, \frac{k}{k_c} - 1.
\end{equation}

For the largest $\epsilon = 0.186$ we have observed for the Ekman 
vortex generating BCII an interesting behavior for very small
$Re < 0.5~(V_g \lesssim 0.3)$. For these parameters the PV~pattern 
did not develop an axially homogeneous wavelength in the spatial 
region where the flow amplitude was indeed spatially uniform. 
Under these circumstances the wavelengths roughly
show a linear variation with $z$ between a larger value (upper small
filled triangles in Fig.~\ref{selectedwavelength}) at the upstream 
end of the bulk amplitude region and a smaller value (lower 
small filled triangles in Fig.~\ref{selectedwavelength}) at 
the downstream end of the bulk region of constant flow amplitude. 
So the Ekman vortex at the inlet
inpedes the free downstream motion of the PV flow
-- it exerts a phase pinning force on the PV~pattern that stretches 
the node distances. The situation is somewhat similar to the explicitly
phase pinning boundary conditions that have been investigated in
Rayleigh-B\'enard convection with through-flow 
\cite{PCGP-Euro-87,MLK-PhysRevA-92}.

\subsubsection{ Finite size effects?} \label{finitesize}

The selected wavelengths in the bulk region are 
practically independent of
the system length as long as a saturation of the Taylor vortices takes
place. To check this we performed a few simulations with aspect ratios
$\Gamma=25$ ($\epsilon=0.114, Re=2$, BCI and BCII) and 
$\Gamma=10$ ($\epsilon=0.114,0.186, Re=1$, BCI and BCII) and 
compared with
the results obtained for our standard length $\Gamma=50$. The change in
the selected wavelength or frequency is less than $0.1\%$ if the aspect 
ratio is halved to $\Gamma=25$. In the system with length $\Gamma=10$ no
bulk region can be observed, therefore the wavelength can not reach a
constant value. The oscillation frequency of the flow pattern is always
independent of the axial position and mainly unchanged; the deviations 
are less than $0.5\%$ even for the very short system of $\Gamma=10$.

We have also investigated the dependence of nonlinear, saturated 
PV~flow on the numerical discretization. These tests show that 
the nonlinear vortex
structures are basically independent of the discretization -- provided 
it is not to coarse -- if one bases the comparison of bifurcated
flow structures obtained with different discretizations on the relative
control parameter $\mu = \frac{T}{T_c} - 1 $ that is influenced via 
$T_c$ by the discretization in question \cite{NonlinearPV}. 

\subsubsection{Comparison with front propagation -- NSE $vs$ GLE}
\label{comparisonfront}

Let us first consider Taylor vortices without through-flow. Then the GLE
(\ref{GLE}) contains no imaginary coefficients $c_i$ and the first-order
spatial derivative is absent since $v_g=0$. In this case the GLE~yields
for propagating fronts \cite{DL-PhysRev-83} as well as for stationary 
patterns in finite systems \cite{CDHS-FluidMech-83}
the critical wave number, $q(z) = 0$, all
over the extension of the pattern -- the boundary condition $A = 0$ 
enforces the collaps of the supercritical band of stable bulk wave 
numbers of {\it nonlinear} vortex patterns to $q=0$. 
This {\it nonlinearly} selected wave number $q=0$ happens to be 
for $Re=0$
the same as the wave number of maximal {\it linear} growth
under a {\it linear} front whose spatiotemporal evolution is governed
by the dispersion $\zeta_{GLE} (Q, \epsilon; Re=0)$ (\ref{OMQBULK})
of the {\it linear} GLE. In the presence of through-flow, however, the 
wave number of maximal growth under the {\it linear} front --
e.g. $q_F$ (\ref{QFLIN}) -- of the GLE differs from the one in the 
{\it nonlinear} bulk far behind the front 
[$(q_b)_F$ (\ref{QBULKFRONT})].

Now, already without through-flow the dispersion relation of the
{\it linear} NSE differs significantly from that of the GLE so that the 
wave number of largest temporal growth under a {\it linear} NSE front 
deviates from the GLE result $q=0$. 
As an aside we mention that a somewhat similar behavior
has also been observed experimentally \cite{AC-PhysLett-83,FS86} and 
numerically for Taylor vortices \cite{LMW-PhysRevA-85,Niklas-un} 
or Rayleigh-B\'enard convection rolls \cite{LMK-PhysRevA-87} which, 
however, 
were located in the nonlinear part of propagating fronts. 
In this context it is useful to keep in mind that the wavelength 
profile $\lambda (z)$ of numerically obtained Taylor vortex fronts
\cite{LMW-PhysRevA-85,Niklas-un} shows an axial variation with a 
characteristic dip in the steepest region of the front similar to the
one of the full line in Fig.~\ref{fouriermodesBSBC}b near $z = 8$.

Now let us consider the through-flow case.
In Fig.~\ref{omvalues} we compare the eigenfrequencies of the
{\it nonlinear} NSE (open symbols) for BCI with those (solid line) 
that would be selected by a {\it linear} front that is stationary 
at the border of the absolutely unstable regime. To that 
end Recktenwald and Dressler
\cite{RDprivate} have determined the dispersion relation 
$\zeta (Q)$ of linear perturbations 
$\sim e^{i[(k_c + Q) z - (\omega_c + \zeta) t]}$ in
the complex $Q$-plane resulting from the NSE with a shooting 
method as described in ref.~\cite{BAC-Pre-93}. The solid line 
in Fig.~\ref{omvalues} represents the frequency 
$\Omega_F = Re(\zeta (Q_s))$ selected by the
stationary front $(v_F = 0)$ of the {\it linear} NSE. Again, as 
discussed for the GLE in Sec.~\ref{patternselectiona} 5, this 
frequency is determined by the saddle $Q_s$ of $\zeta (Q)$ for 
which the temporal growth rate
$Im~\zeta (Q_s) = 0$. This defines the boundary between absolute 
and convective instability. 
With increasing $Re$ the eigenfrequency of the {\it nonlinear} 
NSE approaches the frequency selected by the {\it linear} 
stationary front at the border of the absolutely unstable regime 
as indicated for the
three $\epsilon$-values shown in Fig.~\ref{omvalues}. The front 
frequencies for these $\epsilon$-values are marked by small open 
symbols on the curve $Re(\zeta (Q_s))$ to indicate that the 
eigenfrequencies of the NSE
in the absolutely unstable regime (large open symbols in 
Fig.~\ref{omvalues}) 
do indeed end at the right positions on this curve.

This NSE behavior is
very similar to the GLE behavior (halftone symbols and dashed line).
Note, however, that the linear dispersion relations of NSE and GLE 
are different: the front frequency
$Re(\zeta_{NSE}(Q_s))\simeq 0.0072 Re + 0.0056 Re^3$ 
(solid line in Fig.~\ref{omvalues})
is positive and increases with through-flow while
$Re(\zeta_{GLE}(Q_s))\simeq -0.0002 Re - 0.0062 Re^3$ (dashed line) 
is negative and decreases with through-flow. This difference in the
dispersion relations seems to be the major cause for the 
differences in the selected patterns.

\subsubsection{Comparison with experiments} 

In early experiments by Snyder \cite{S-Lond-61,S-Ann-65} or Takeuchi 
and Jankowski \cite{TJ-FluidMech-81} the distinction between absolute 
and convective instability was not yet established. 
PV patterns were observed with wavelengths below the critical one 
for  Reynolds numbers $Re<10$ \cite{S-Lond-61,S-Ann-65}. However,
a pattern selection was not reported. The 
wavelengths were seen to vary up to $7\%$ from run to run 
for the same control parameter combinations
and in some cases $5 \%$ over the spatial extension. Furthermore, 
the vortex spacings 
did not change after the occurrence of PV flow 
when increasing the relative Taylor number up to $\epsilon=0.15$ 
\cite{S-Ann-65}. 

Takeuchi and Jankowski \cite{TJ-FluidMech-81} obtained wavelengths 
below the critical one for example with a deviation of 
$\approx - 2.5 \%$ for 
$Re=9.4,\epsilon \approx 0.06, \eta=0.5$ that is in the same order of 
magnitude of our NSE results in the absolute 
instability regime. 

The experimental results of Tsameret and
Steinberg \cite{TS-Euro-91,TS1-93} exhibit a unique selection, similar 
to the results presented here.
Their wavelengths \cite{TS-Euro-91} are always less than the critical 
one for axial boundaries that enforce stationary Ekman 
vortex flow.  However, in contrast to our results they observed no 
$\epsilon$ dependence of the selected wavelengths but an  
Re dependence that is much stronger than ours leading to a relative
deviation from $\lambda_c$ of up to $-10\%$ at $Re \approx 3$ for 
$\eta=0.707$ \cite{TS-Euro-91}. 
In addition their phase 
velocity of PV flow, $v_p=1.055 Re$ \cite{TS-Euro-91}, is much smaller 
than ours, which is close to the critical one. All wavelengths observed 
for through-flow rates above $Re=1$ belong to noise 
sustained structures in the convectively unstable region \cite{TS1-93},
which leads to deviations from $\lambda_c$ of about $3\%$ 
when one limits the comparison to the regime of absolute instability.
Since the $\epsilon$-values at which the measurements of
\cite{TS-Euro-91,TS1-93} were performed are not given we have estimated
an $\epsilon$-value of $\approx 0.025$ based on the convective 
instability border at $Re=1$. For this $\epsilon$ the above mentioned 
$3\%$ deviation of $\lambda$ from $\lambda_c$ is an order of
magnitude larger than our NSE results or the results  
(\ref{QCCONV}) of the GLE.

On the other hand, the experimental wave number found in \cite{TS3-93}
for a known, published $\epsilon \approx 0.38$ and $Re$ up to 4 can 
be compared directly with our results to check the predicted 
$\epsilon$ and $Re$ dependence. In this case all experimental
wavelengths \cite{TS3-93} decrease up to $10 \%$ below the 
critical one at $Re \approx 3.7$ whereas we would expect from our 
simulations and from \cite{MLK-Euro-89,MLK-PhysRevA-92} less than 
half the deviation from $\lambda_c$. 

%%%%%%%%%%%%%%%%%%%%%%%%%%%%%%%%%%%%%%%%%%%%%%%%%%%%%%%%%%%%%%%%%%%%%%%
%
%  SECTION V: Conclusion
%
%%%%%%%%%%%%%%%%%%%%%%%%%%%%%%%%%%%%%%%%%%%%%%%%%%%%%%%%%%%%%%%%%%%%%%%

\section{Conclusion}

We have investigated propagating vortex structures in an axial 
flow for different realistic axial boundary conditions in systems 
of finite length. Within the subregion of absolute instabilty 
a unique pattern
selection is observed. The selected PV flow structures are
independent of parameter history, initial conditions, and
system's length (provide it is large enough to allow for a 
nonlinear, saturated, homogeneous bulk region).
But they depend on the axial boundary conditions. 
For conditions enforcing 
the basic state at the boundaries as well as for conditions that 
enforce stationary Ekman vortex flow near the boundaries the PV 
pattern is suppressed at the inlet and outlet of the annulus. 
Then one observes a
characteristic stationary axial intensity and wave number profile,
whereas the
oscillation frequency of the pattern is constant all over the system. 
The analysis of the appropriate Ginzburg-Landau equation for 
PV flow shows that the selected frequency of the pattern oscillation 
is the eigenvalue of a nonlinear eigenvalue problem for which the axial
variation of the corresponding eigenfunction is smooth and as small as
possible. The eigenfunction is the complex pattern amplitude 
that characterizes the intensity and wave number profiles of the PV
structure. The GLE intensity profiles agree well with those of the NSE.
However, there are characteristic differences in the selected 
frequencies and wave number
profiles of the full NSE and the GLE approximation. They are identified 
to be mainly caused by the different dispersion relations of the
equations. Approaching the border between absolute and convective
instability the eigenvalue problem becomes effectively linear and the
pattern selection mechanism becomes that one of linear front 
propagation. 

\acknowledgments

This work was supported by the Deutsche Forschungsgemeinschaft 
and the Stiftung Volkswagenwerk.

%%%%%%%%%%%%%%%%%%%%%%%%%%%%%%%%%%%%%%%%%%%%%%%%%%%%%%%%%%%%%%%%%%%%%%%
%
%  References
%
%%%%%%%%%%%%%%%%%%%%%%%%%%%%%%%%%%%%%%%%%%%%%%%%%%%%%%%%%%%%%%%%%%%%%%%

%%%%%%%%%%%%%%%%%%%%%%%%%%%%%%%%%%%%%%%%%%%%%%%%%%%%%%%%%%%%%%%%%%%%%%%
%
%  TABLE I
%
%%%%%%%%%%%%%%%%%%%%%%%%%%%%%%%%%%%%%%%%%%%%%%%%%%%%%%%%%%%%%%%%%%%%%%%

\begin{table}

\caption{Bold-faced numbers denote critical values and 
coefficients that are appropriate for the finite-differences code 
with spatial grid size $0.05$ used in
our simulations of the NSE and that have been used for scaling the NSE 
results. The methods for determining these numbers are described in 
Sec.~II C.
The imaginary parts $c_1$, $c_2$, $c_3$ are taken from ref.~[27].
$\gamma$ depends on the normalization of the linear radial 
eigenfunction $\hat{u}\left(r\right)$ which was chosen to be
$|\hat{u}\left(r_1+0.5\right)|=15.2$.}

\begin{minipage} [b]{8.5cm}
\begin{tabular}{cccccc} 
\multicolumn{6}{c}
{$a = a_0 \left[1+\frac{a_2}{\left|a_2 \right|}
\left(Re/a_2 \right)^2 \right]$} \\ 
& $T_c$ & $k_c$ & $\xi^2_0$ & $\tau_0$ & $\gamma$ \\ \hline
$a_0$ & {\bf 2420.23} & {\bf 3.1305} & {\bf 0.144} & {\bf 0.0762} & 
{\bf 8.06} \\ 
$a_2$  & {\bf 138.62} & {\bf 252.96} & {\bf -27.23} & {\bf -49.68} & 
{\bf -12.33} \\ \hline \hline
\multicolumn{6}{c}
{$a = a_1 Re \left[ 1 + \frac{a_3}{\left|a_3 \right|}
\left( Re/a_3 \right)^2 \right]$}\\
& $v_g$ & $\omega_c$ & $c_0$ & $c_1$ & $c_2$ \\ \hline
$a_1$ & {\bf 1.20}  & {\bf 3.647} &  {\em 1/138} & {\em 1/40.7} & 
{\em 1/287} \\ 
$a_3$ & -- & -- & {\em -51.8} & {\em 70.5} & {\em -97.8}\\ 
\end{tabular}

\end{minipage}

\end{table}

%%%%%%%%%%%%%%%%%%%%%%%%%%%%%%%%%%%%%%%%%%%%%%%%%%%%%%%%%%%%%%%%%%%%%%%
%
%  CAPTIONS
%
%%%%%%%%%%%%%%%%%%%%%%%%%%%%%%%%%%%%%%%%%%%%%%%%%%%%%%%%%%%%%%%%%%%%%%%

% fig1 
%%%%%%%%%%%%%%%%%%%%%%%%%%%%%%%%%%%%%%%%%%%%%%%%%%%%%%%%%%%%%%%%%%%%%%%

\begin{figure} \caption[]
{Stability domains of the basic flow
state (\ref{BFLOW}) in the plane of control parameters. Numerical
simulations have been performed for the parameters marked by circles
($\epsilon=0.0288$), diamonds ($\epsilon=0.114$), and triangles
($\epsilon=0.186$). 
Dashed line is the critical threshold for onset of
extended PV flow and full line the boundary (\ref{MUCCONV}) between 
absolute
and convective instability. The parameters entering (\ref{MUCCONV}) --
cf.~Table I -- characterize the finite-differences version of the 
NSE. They have been obtained with the methods described in Sec.~II C.}
\label{controlparameterplane}
\end{figure}

%fig2 
%%%%%%%%%%%%%%%%%%%%%%%%%%%%%%%%%%%%%%%%%%%%%%%%%%%%%%%%%%%%%%%%%%%%%%%

\begin{figure} \caption[] 
{Spatiotemporal structure of PV flow. Thin lines show vertically 
displaced snapshots of the axial
velocity field $w(r_1+0.225,z;t)$ at successive, equidistantly 
spaced times. Thick lines show the stationary envelopes. The BCI at
inlet and outlet suppresses any vortex flow there, while the BCII  
suppresses PV flow but induces {\it stationary} Ekman vortices.
Parameters are $\epsilon=0.114$ and $Re=2.5$.}
\label{hiddenlineplot}
\end{figure}

%fig3
%%%%%%%%%%%%%%%%%%%%%%%%%%%%%%%%%%%%%%%%%%%%%%%%%%%%%%%%%%%%%%%%%%%%%%%

\begin{figure} \caption[]
{Scaled growth length $L$ (\ref{L}) of
PV structures vs scaled group velocity $V_g$ (\ref{Vg}). 
Symbols represent lengths
obtained from the NSE for different combinations 
(cf.~Fig.~\ref{controlparameterplane})
of $Re$ and $\epsilon=0.0288$ (circles), $\epsilon=0.114$ (diamonds), 
and $\epsilon=0.186$ (triangles). Boundary conditions are BCI for open
symbols and BCII for filled ones. The line shows the scaling 
behavior of the GLE subject to BCI.}
\label{Lscaled} 
\end{figure}

%fig4
%%%%%%%%%%%%%%%%%%%%%%%%%%%%%%%%%%%%%%%%%%%%%%%%%%%%%%%%%%%%%%%%%%%%%%%

\begin{figure} \caption[]
{Structure of PV flow selected for BCI at inlet and outlet. Shown 
are the axial variations of the {\it temporal} Fourier modes (a) and 
of the 
wavelength (b) of $w (r_{1}+0.225,z;t)$. Full lines result from
numerical simulations of the full NSE. Dashed lines come from the
GLE. Parameters are $\epsilon=0.114$ and $Re=2.5$.}
\label{fouriermodesBSBC}
\end{figure}

%fig5
%%%%%%%%%%%%%%%%%%%%%%%%%%%%%%%%%%%%%%%%%%%%%%%%%%%%%%%%%%%%%%%%%%%%%%%

\begin{figure} \caption[]
{Axial structure of vortex patterns subject to BCII. The zeroth
temporal Fourier mode $w_0$ (thick lines) of the axial velocity
field $w(r_1+0.225,z;t)$
reflects the stationary Ekman vortices. The modes $w_n$ 
with $n>0$ (thin lines) characterize the oscillating PV pattern.
Parameters are $\epsilon=0.114$ and $Re=1.0$ (a), $2.5$ (b),
$2.7$ (c).}
\label{zerothfouriermodeEVBC}
\end{figure}

%fig6
%%%%%%%%%%%%%%%%%%%%%%%%%%%%%%%%%%%%%%%%%%%%%%%%%%%%%%%%%%%%%%%%%%%%%%%

\begin{figure} \caption[]
{Selected wavelengths in the bulk region of PV stuctures vs scaled 
group velocity $V_g$ (\ref{Vg}).
Open (halftone) symbols refer to the NSE (GLE) subject to BCI at
inlet and outlet. Filled symbols refer to the NSE subject to BCII
at inlet and outlet. Parameters are
$\epsilon=0.0288$ (circles), $\epsilon=0.114$ (diamonds), and
$\epsilon=0.186$ (triangles).
The small filled triangles are explained in 
Sec.~\ref{patternselectionb}2.}
\label{selectedwavelength}
\end{figure}

%fig7
%%%%%%%%%%%%%%%%%%%%%%%%%%%%%%%%%%%%%%%%%%%%%%%%%%%%%%%%%%%%%%%%%%%%%%%

\begin{figure} \caption[]
{Selected oscillation frequency of PV flow obtained with
BCI from the NSE 
(open symbols) and GLE (halftone symbols) vs Reynolds number. 
The full (dashed) line is the front frequency 
$\Omega_F = Re(\zeta(Q_s))$ 
of the linearized NSE (GLE) at the border between absolute and 
convective instability. Parameters are $\epsilon=0.0288$ (circles),
$\epsilon=0.114$ (diamonds), and $\epsilon=0.186$ (triangles).
The small symbols marking $\Omega_F$ for these three $\epsilon$ values
show that the eigenfrequencies of the nonlinear equations in the 
absolutely unstable regime approach the $\Omega_F$ limit curve at 
the right places. }
\label{omvalues}
\end{figure}

%fig8
%%%%%%%%%%%%%%%%%%%%%%%%%%%%%%%%%%%%%%%%%%%%%%%%%%%%%%%%%%%%%%%%%%%%%%%

\begin{figure} \caption[]
{Frequency shift $\omega - \omega_c$ of PV flow vs scaled group 
velocity $V_g$ (\ref{Vg}) for BCI. Both, NSE (open symbols) and GLE 
(halftone symbols) results are scaled 
by the corresponding GLE frequency $\Omega_{conv}^c$ (\ref{OMCCONV})
at the convective instability border for the respective
parameters $\epsilon=0.0288$ (circles), $\epsilon=0.114$ (diamonds), 
and
$\epsilon=0.186$ (triangles). The small symbols have the same meaning 
as explained in Fig.~\ref{omvalues}.}
\label{scaledom}
\end{figure}


\begin{thebibliography}{99}
\bibitem{NonlinearPV}
Using discretizations of $n=15$, $20$, and
$30$ lattice points per unit length we observe a systematic deviation in
the selected wave numbers of the nonlinear PV patterns. For $T=2870$,
$Re=0.75$ and the boundary condition BCII (Sec.~\ref{EVBC}) the selected
wavelength of the saturated PV~flow is $\lambda=2.102,2.089,2.076$ for
$n=15,20,30$, respectively. A coarser grid provides a smaller
$T_c(Re)$ and a bigger $\lambda$ which is in agreement with
Fig.~\ref{selectedwavelength}. In the case of BCI (Sec.~\ref{BSBC}) the
coarser grid decreases the wavelength $\lambda=1.949,1.952,1.953$ for
$n=15,20,30$ at $T=2870$ and $Re=3.1$ also in agreement with 
Fig.~\ref{selectedwavelength}. By using the relative control parameter 
$\mu$ (2.9a) with the appropriate $T_c (Re)$ these systematic 
deviations practically disappear.
\end{thebibliography}
\end{document}